\documentclass[sigconf,bookmarksnumbered,unicode,hidelinks]{acmart}
\usepackage{algorithm}

\usepackage{algorithmicx}
\usepackage{algpseudocode}
\usepackage{tikz}
\usetikzlibrary{fit}
\usetikzlibrary{positioning}
\usepackage{graphicx}
\usepackage{booktabs}
\usepackage{multirow}
\usepackage{array}
\newcolumntype{M}[1]{>{\centering\arraybackslash}m{#1}}
\usepackage{placeins}
\FloatBarrier

\AtBeginDocument{%
  }

\setcopyright{acmlicensed}
\copyrightyear{2025}
\acmYear{2025}
\acmDOI{XXXXXXX.XXXXXXX}
\acmConference[QAI-NLP 2025]{Quantum AI and NLP Conference 2025}{August 6--8, 2025}{Bloomington, IN, USA}

\acmISBN{}

\begin{document}

\title{Enhancing Interpretability of Quantum-Assisted Blockchain Clustering via AI Agent-Based Qualitative Analysis}

\author{Yun-Cheng Tsai}
\authornote{Both authors contributed equally to this research.}
\email{pecu@ntnu.edu.tw}
\author{Yen-Ku Liu}
\authornotemark[1]
\email{41171123h@ntnu.edu.tw}
\affiliation{%
  \institution{National Taiwan Normal University}
  \city{Taipei}
  \country{Taiwan}
}

\author{Samuel Yen-Chi Chen}
\affiliation{%
  \institution{Wells Fargo}
  \city{New York}
  \country{USA}}
\email{ycchen1989@ieee.org}

\renewcommand{\shortauthors}{Y.C. Tsai et al.}

\begin{abstract}
Blockchain transaction data is inherently high-dimensional, noisy, and entangled, posing substantial challenges for traditional clustering algorithms. While quantum-enhanced clustering models have demonstrated promising performance gains, their interpretability remains limited, restricting their application in sensitive domains such as financial fraud detection and blockchain governance. To address this gap, we propose a two-stage analysis framework that synergistically combines quantitative clustering evaluation with AI Agent-assisted qualitative interpretation. In the first stage, we employ classical clustering methods and evaluation metrics—including the Silhouette Score, Davies-Bouldin Index, and Calinski-Harabasz Index—to determine the optimal cluster count and baseline partition quality. In the second stage, we integrate an AI Agent to generate human-readable, semantic explanations of clustering results, identifying intra-cluster characteristics and inter-cluster relationships. Our experiments reveal that while fully trained Quantum Neural Networks (QNN) outperform random Quantum Features (QF) in quantitative metrics, the AI Agent further uncovers nuanced differences between these methods, notably exposing the singleton cluster phenomenon in QNN-driven models. The consolidated insights from both stages consistently endorse the three-cluster configuration, demonstrating the practical value of our hybrid approach. This work advances the interpretability frontier in quantum-assisted blockchain analytics and lays the groundwork for future autonomous AI-orchestrated clustering frameworks.
\end{abstract}

\ccsdesc[500]{Computing methodologies~Machine learning}
\ccsdesc[500]{Computing methodologies~Modeling and simulation}
\ccsdesc[500]{Information systems~Information integration}

\keywords{Machine Learning, Blockchain Analytics, AI Agent-based Explainability, Hybrid Quantum-Classical Clustering, Explainable Artificial Intelligence (XAI)}


\maketitle

\section{Introduction}

Blockchain analytics has emerged as a critical field for understanding decentralized economic activities and ensuring network security. However, the intrinsic properties of blockchain data---high dimensionality, temporal entanglement, and transactional noise---pose significant challenges for conventional clustering methods, often leading to suboptimal performance and opaque results \cite{saputra2024blockchain, ahmed2020kmeans}. Recent advancements in quantum machine learning (QML) offer promising avenues to overcome these hurdles by enhancing feature representations through quantum state encoding \cite{biamonte2017quantum, havlivcek2019supervised}. Yet, these quantum-assisted models primarily optimize for clustering performance metrics and lack transparency, which undermines their adoption in sensitive applications such as anti-money laundering (AML), fraud detection, and decentralized finance (DeFi) ecosystem monitoring \cite{das2020xai, miller2019explanation}.

To bridge this gap, we propose a \textbf{two-stage analysis framework} that unites the rigor of quantitative evaluation with the interpretability of AI Agent-driven qualitative analysis. In the first stage, we systematically evaluate clustering quality across varying cluster counts ($K=2$ to $K=6$), utilizing classical metrics like Silhouette Score, Davies-Bouldin Index, and Calinski-Harabasz Index to identify the most promising partitioning strategies \cite{ashari2023analysis}. This phase ensures an objective selection of baseline clustering configurations, effectively narrowing the search space for subsequent analysis.

The second stage employs an AI Agent, leveraging large language models to perform semantic analysis on the clustering outputs. The Agent interprets intra-cluster characteristics, reveals cross-cluster relationships, and identifies potential anomalies---providing critical explanations beyond numerical metrics \cite{nazar2021hci}. Furthermore, the AI Agent is more than a passive observer; it evolves toward an active role by proposing adjustments to cluster configurations and offering strategic insights into clustering parameter design.

Our key contributions are fourfold:

\begin{enumerate}
    \item \textbf{Two-stage analysis framework:} We establish a novel hybrid evaluation pipeline combining quantitative clustering assessments with AI Agent-driven qualitative insights.
    \item \textbf{Enhanced interpretability:} Our approach improves the explainability of quantum-assisted clustering models, particularly in blockchain applications where actionable insights are paramount.
    \item \textbf{AI Agent as an active participant:} We advance the role of AI Agents from passive explainers to proactive collaborators, capable of suggesting cluster configurations and detecting anomalies.
    \item \textbf{Empirical validation in blockchain scenarios:} Through comprehensive experiments on carbon credit transaction data, we demonstrate the effectiveness of our method, with both quantitative and qualitative analyses converging on a three-cluster solution as the optimal configuration.
\end{enumerate}

This study addresses the long-standing interpretability challenges in quantum machine learning. It opens new directions for autonomous AI-driven analytics, paving the way toward fully automated, explainable clustering systems in complex data environments.

\section{Related Work}

\subsection{Quantum Feature Engineering in Blockchain Analytics}
Quantum-enhanced feature engineering has emerged as a promising solution to tackle high-dimensional and entangled blockchain data~\cite{wittek2014quantum}. Quantum Neural Networks (QNNs) utilize quantum properties such as superposition and entanglement to project classical data into high-dimensional Hilbert spaces, revealing latent patterns challenging to capture with classical methods~\cite{cerezo2021variational}.

\subsection{Hybrid Classical-Quantum Clustering Approaches}
Previous research has explored hybrid strategies, combining quantum-generated features with classical clustering algorithms~\cite{havlivcek2019supervised}. Untrained quantum circuits (Random QNN) have been used to generate rich feature sets, demonstrating significant improvements in clustering tasks without extensive training overhead~\cite{preskill2018quantum}.

\subsection{Explainable AI (XAI) in Quantum Machine Learning}
Explainable AI aims to make black-box models more transparent. In quantum machine learning, interpretability remains an under-explored frontier~\cite{nazar2021systematic,lin2024quantum}. Our work contributes to this field by integrating an AI Agent to transform complex quantum clustering outcomes into digestible insights~\cite{das2020opportunities}.

\subsection{AI Agents for Model Interpretation}
AI Agents, particularly those leveraging natural language generation and rule-based reasoning, have effectively summarized complex model behaviors~\cite{miller2019explanation}. We enable automated, interpretable reporting of quantum-enhanced clustering results by deploying such an agent in our study.

\section{Methodology}
\begin{figure*}[ht]
\centering
\begin{tikzpicture}[node distance=1.8cm and 3.5cm, auto]

    \tikzstyle{startstop} = [rectangle, rounded corners, minimum width=3cm, minimum height=0.7cm, text centered, draw=black, fill=red!30]
    \tikzstyle{process} = [rectangle, minimum width=3.5cm, minimum height=0.7cm, text centered, draw=black, fill=orange!30]
    \tikzstyle{arrow} = [thick,->,>=stealth]

    \node (start) [startstop] {Start};
    \node (preproc) [process, below of=start] {Data Collection};

    \node (qrand) [process, below left of=preproc, xshift=-3.5cm, yshift=-0.6cm] {Quantum Random Feature Extraction};
    \node (qtrain) [process, below right of=preproc, xshift=3.5cm, yshift=-0.6cm] {Train QNN (Composite Loss)};

    \node (comb2) [process, below of=qrand] {Hybrid Feature Set};
    \node (qclust) [process, below of=qtrain] {Quantum Clustering};

    \node (kmeans2) [process, below of=comb2] {K-Means};

    \node (eval) [process, below of=preproc, yshift=-5.2cm] {Silhouette Score \& Davies-Bouldin \& Calinski-Harabasz Evaluation};
    \node (result) [startstop, below of=eval] {Qualitative analysis phase};

    \draw [arrow] (start) -- (preproc);
    \draw [arrow] (preproc) -- (qrand);
    \draw [arrow] (preproc) -- (qtrain);

    \draw [arrow] (qrand) -- (comb2);
    \draw [arrow] (comb2) -- (kmeans2);
    \draw [arrow] (kmeans2) -- (eval);

    \draw [arrow] (qtrain) -- (qclust);
    \draw [arrow] (qclust) -- (eval);

    \draw [arrow] (eval) -- (result);

\end{tikzpicture}
\caption{Overview of the experimental design for quantum-enhanced clustering.}\Description{This figure shows the experimental design for quantum-enhanced clustering.}
\label{fig:flowchart}
\end{figure*}
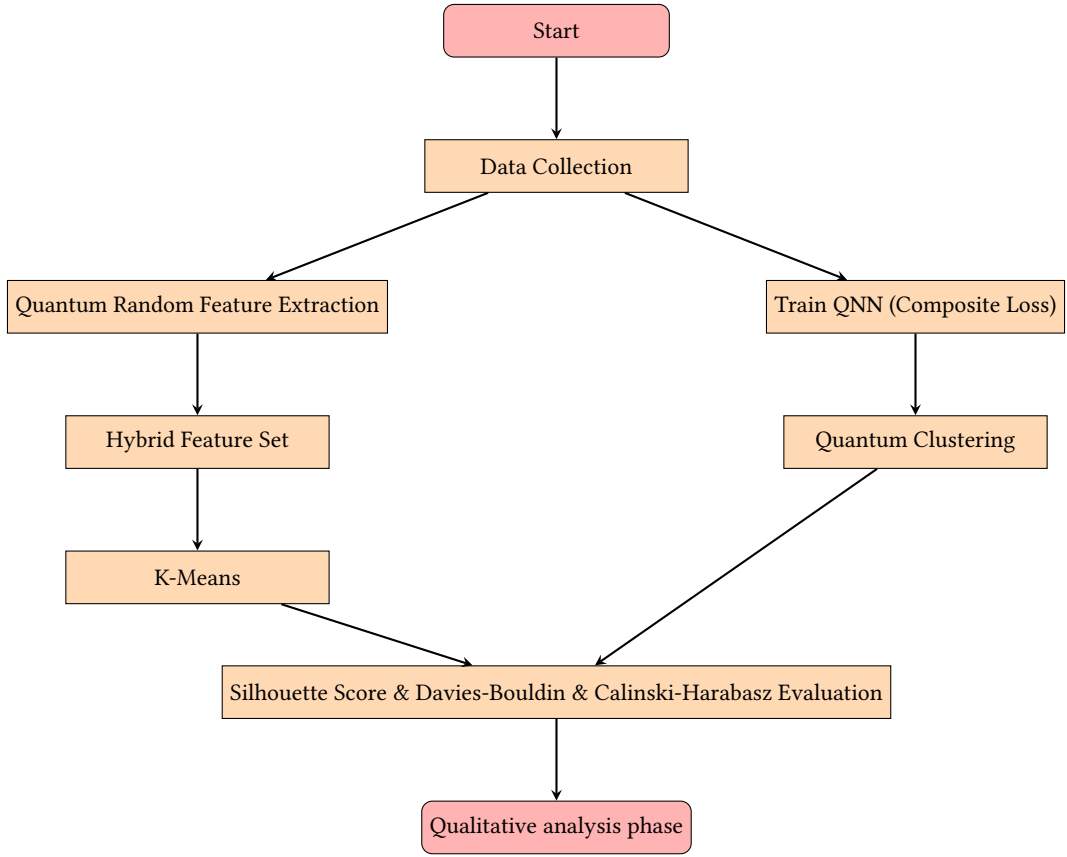

This section presents our two-stage analysis framework, designed to enhance blockchain transaction data's clustering quality and interpretability. The proposed approach integrates rigorous quantitative evaluation with AI Agent-assisted qualitative analysis, enabling a holistic understanding of clustering behaviors. The overall workflow is illustrated in Figure~\ref{fig:flowchart}.

\subsection{Stage 1: Quantitative Clustering Evaluation}

\subsubsection{Objective}
The first stage aims to establish an objective baseline of clustering performance using established evaluation metrics. We systematically evaluate clustering quality across varying cluster counts ($K=2$ to $K=6$) to identify optimal configurations for deeper analysis.

\subsubsection{Dataset and Preprocessing}
We employ blockchain transaction data from the MCO2 token ecosystem on Ethereum, focusing on carbon credit trading activities involving ten major nodes~\cite{etherscanMCO2}. The dataset captures essential attributes such as block numbers, timestamps, sender and receiver addresses, token details, transaction values, and gas prices.

A detailed summary of the dataset attributes and preprocessing techniques is provided in Table~\ref{tab:dataset_summary}.

\begin{table}[htbp]
\centering
\caption{Dataset Attribute Summary and Preprocessing Techniques}
\label{tab:dataset_summary}
\begin{tabular}{|l|l|l|}
\hline
\textbf{Attribute} & \textbf{Description} & \textbf{Preprocessing} \\
\hline
block\_number & Block height & Numerical, no scaling \\
transaction\_hash & Transaction ID & Ignored (non-informative) \\
timestamp & Transaction time & Cyclical encoding (sin/cos) \\
from\_address & Sender address & Label encoding \\
to\_address & Receiver address & Label encoding \\
token\_name & Token name & Label encoding \\
token\_symbol & Token symbol & Label encoding \\
token\_value & Amount transferred & Robust scaling \\
gas\_price & Transaction fee & Robust scaling \\
\hline
\end{tabular}
\end{table}

Categorical attributes are transformed through label encoding, while numerical variables like \texttt{token\_value} and \texttt{gas\_price} undergo robust scaling to mitigate outliers. The \texttt{timestamp} is converted into cyclical features (sine and cosine) to capture temporal patterns in transaction behaviors.

This structured preprocessing pipeline ensures the feature set is optimized for quantum feature extraction and clustering.

\subsubsection{Quantum Feature Extraction Strategies}
We propose two distinct quantum-enhanced feature extraction strategies, each built upon the Quantum-Train (QT) paradigm, which utilizes a QNN with its exponential number of amplitudes to generate the parameters of a classical neural network \cite{liu2024quantum}.

\begin{itemize}
    \item \textbf{Random Quantum Features (QF):} 
    The features are extracted from classical neural networks whose parameters are derived from randomly initialized, untrained QNNs, thereby providing diverse and computationally efficient representations.
    \item \textbf{Fully Trained Quantum Neural Network (QNN):} 
    In contrast, the second strategy trains the QNN end-to-end using the SwAV self-supervised loss to generate parameters for the feature transformation network, thereby optimizing the resulting embeddings for clustering performance.
\end{itemize}
Both strategies combine quantum features with preprocessed classical features to form hybrid representations. We employ the K-Means algorithm for clustering due to its effectiveness in unsupervised learning tasks~\cite{prati2004survey}.

\subsubsection{Fully Trained Quantum Neural Network (QNN)}
\begin{figure}[htbp]
\centering
\includegraphics[width=1\columnwidth]{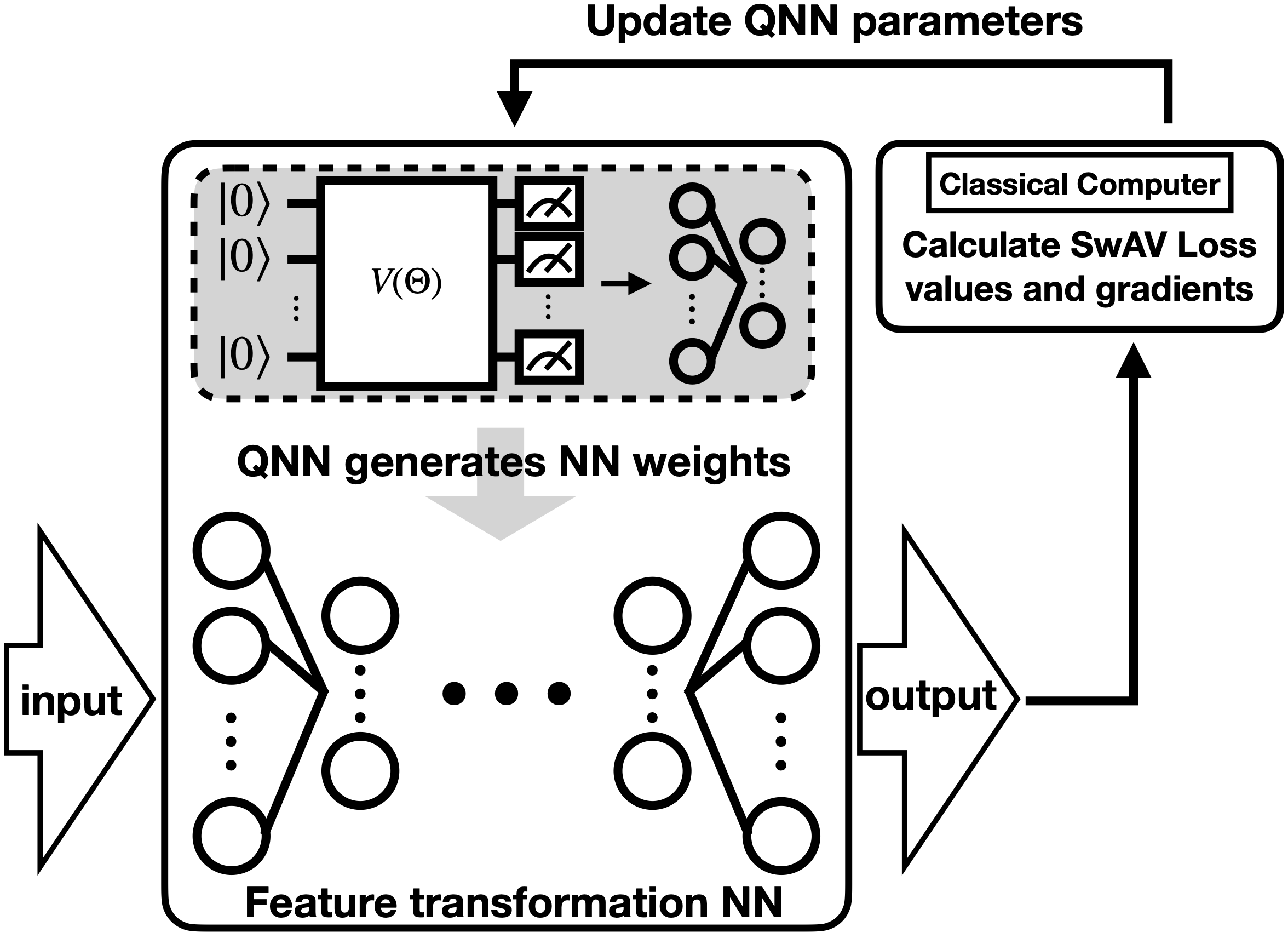}
\caption{Trained QNN for generating weights for feature transformation NN.}
\Description{Diagram showing the structure of a trained quantum neural network used to generate weights for a feature transformation neural network.}
\label{fig:trained_QNN}
\end{figure}

In this approach, we further investigate whether optimizing the QNN parameters can enhance clustering performance. Under the hybrid quantum-classical paradigm, the QNN is integrated with a SwAV loss module, forming a unified model. The entire hybrid model is trained end-to-end, utilizing standard backpropagation for optimization. This scheme is illustrated in Figure~\ref{fig:trained_QNN}.
The loss function is defined as:
\begin{equation}
L_{\text{SwAV}} = \text{KL}\left(\text{log1p}(P_1) \,\|\, P_2\right),
\end{equation}
where \(P_1\) and \(P_2\) are the probability distributions obtained from the QNN outputs with temperature scaling and label smoothing. This loss guides the QNN in learning robust feature representations that are subsequently used for K-means clustering. Training is performed using the Adam optimizer with gradient clipping to mitigate gradient explosion. The Algorithm~\ref{alg:qnn_swav} outlines the fully quantum clustering approach.

\begin{algorithm}[htbp]
\caption{Quantum Neural Network Training with SwAV Loss}
\label{alg:qnn_swav}
\raggedright
\begin{algorithmic}[1]
\For{each $num\_prototypes$ in $prototype\_range$}
    \State Create output directories for current $num\_prototypes$
    \For{each $quantum\_depth$ in $quantum\_depth\_range$}
        \State Initialize base neural network model
        \State Initialize QNN with given $quantum\_depth$ using the base model
        \State Initialize the SwAV loss module with current $num\_prototypes$
        \State Set optimizer with parameters from QNN and SwAV loss

        \For{each epoch in $num\_epochs$}
            \For{each batch in $scaled\_features$}
                \State Set QNN to training mode
                \State $quantum\_output \gets QNN(batch)$
                \State $aug_1 \gets quantum\_output + \mathcal{N}(0, \sigma^2)$
                \State $aug_2 \gets quantum\_output + \mathcal{N}(0, \sigma^2)$
                \State $loss \gets SwAV\_loss(aug_1, aug_2)$
                \State optimizer.zero\_grad()
                \State Backpropagate $loss$
                \State Apply gradient clipping
                \State optimizer.step()
            \EndFor

            \State Set QNN to evaluation mode
            \State Initialize empty list $quantum\_features$
            \ForAll{each batch in $scaled\_features$}
                \State $quantum\_output \gets QNN(batch)$ without gradient computation
                \State Append $quantum\_output$ to $quantum\_features$
            \EndFor
            \State Concatenate $quantum\_features$
        \EndFor
    \EndFor
\EndFor
\end{algorithmic}
\end{algorithm}

\subsubsection{Clustering Evaluation Metrics}

Clustering performance is assessed using three established internal metrics:
\begin{itemize}
    \item \textbf{Silhouette Score}~\cite{rousseeuw1987silhouettes}: Measures cohesion within clusters and separation between clusters (higher is better).
    \item \textbf{Davies-Bouldin Index}~\cite{davies1979cluster}: Evaluates cluster similarity; lower values indicate better separation.
    \item \textbf{Calinski-Harabasz Index}~\cite{calinski1974dendrite}: Compares inter-cluster dispersion to intra-cluster dispersion (higher is better).
\end{itemize}

Optimal configurations are selected based on these metrics for subsequent qualitative analysis.

Performance comparisons of QF and QNN methods across varying $K$ values are presented in Table~\ref{tab:best_results}.

\subsection{Stage 2: AI Agent-Assisted Qualitative Analysis}

\subsubsection{Objective}

While quantitative metrics provide critical performance benchmarks, they lack semantic insights into clustering behaviors. Stage 2 addresses this gap by employing an AI Agent to deliver human-interpretable explanations of the clustering outcomes.

\subsubsection{AI Agent Architecture}

\begin{figure*}[!t]
    \centering
    \includegraphics[width=\textwidth]{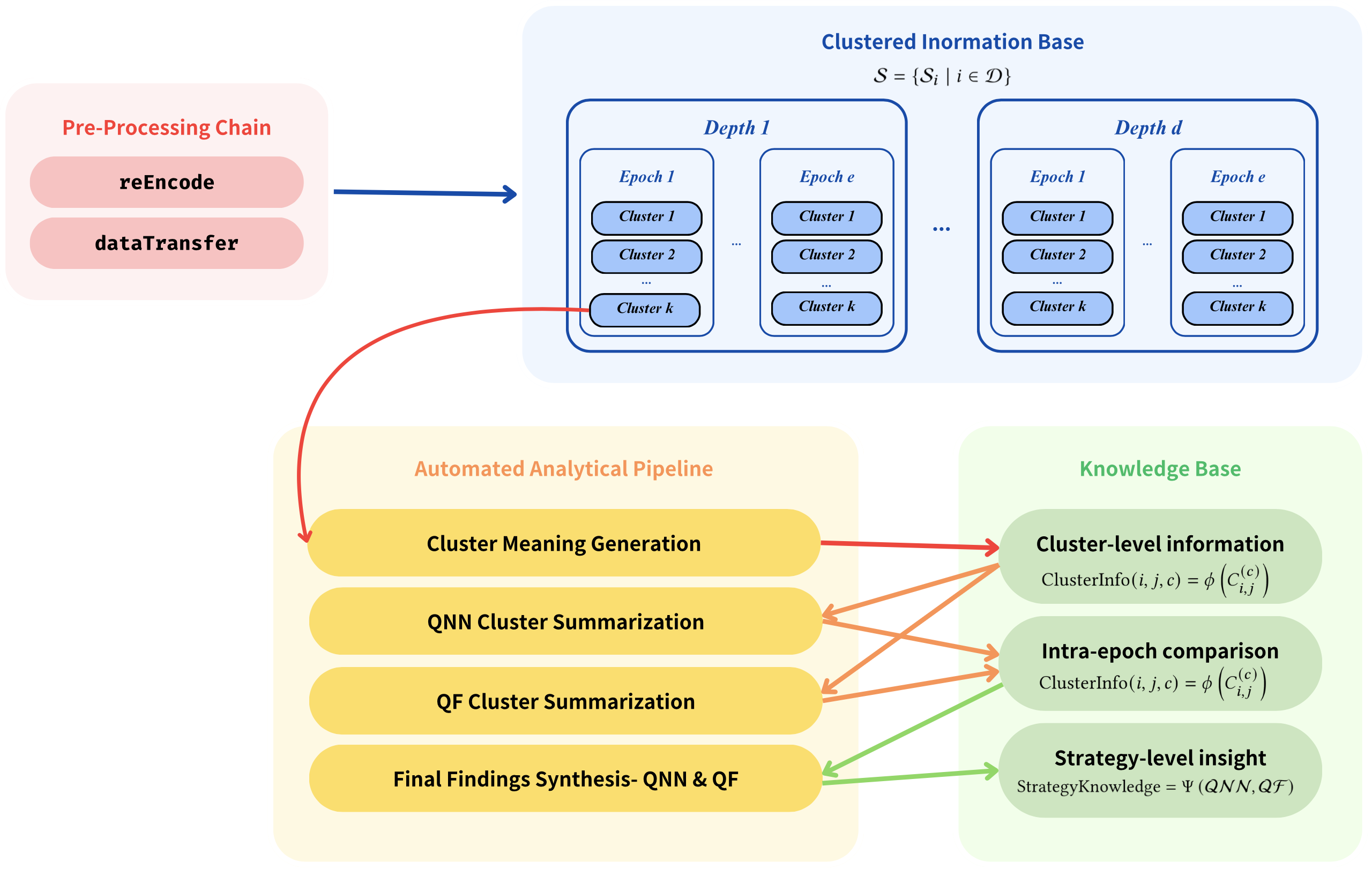}
    \caption{Architecture of the AI Agent used in the qualitative analysis phase.}
    \label{fig:architecture-3}
    \Description{Architecture diagram of the AI Agent used in the qualitative analysis phase.}
\end{figure*}

Our AI Agent consists of three interconnected modules (Figure~\ref{fig:architecture-3}):

\begin{itemize}
    \item \textbf{Pre-processing Chain:} Re-encodes blockchain records into compact, semantically meaningful representations for efficient processing.
    \item \textbf{Automated Analytical Pipeline:} Generates semantic descriptions for each cluster, compares QF and QNN strategies, and synthesizes insights.
    \item \textbf{Knowledge Base:} Maintains memory of semantic summaries, intra-epoch comparisons, and global strategy insights.
\end{itemize}

The AI Agent is designed to answer critical interpretability questions:
\begin{itemize}
    \item What defines the transactional characteristics of each cluster?
    \item Are specific cluster counts overly simplistic or unnecessarily complex?
    \item Should adjustments to clustering configurations be considered?
\end{itemize}

\subsubsection{Pre-processing Chain}

The first step in the AI Agent workflow is transforming raw blockchain records into a compact and semantically meaningful representation. Specifically, we perform one-to-one re-encoding of hashed values, such as transaction IDs, sender addresses, and receiver addresses, to optimize language model efficiency and reduce token consumption.

Beyond simple re-encoding, we architect a shared hierarchical structure, referred to as the \textit{Clustered Information Base}, to organize the processed data systematically. This structure supports both the QNN and QF modules, enabling time-aware and depth-aware partitioning of information. Such structuring ensures scalability and facilitates interpretable downstream analysis.

Formally, the clustered representation is defined as:
\begin{equation}
\mathcal{S} = \left\{ \mathcal{S}_i \mid i \in \mathcal{D} \right\}
\end{equation}

where $\mathcal{D} = \{1, 2, \dots, d\}$ represents depth levels, and each depth contains epochs:
\begin{equation}
\mathcal{S}_i = \left\{ \mathcal{E}_{i,j} \mid j \in \mathcal{E} \right\}
\end{equation}

Further, each epoch is partitioned into clusters:
\begin{equation}
\mathcal{E}_{i,j} = \left\{ C_{i,j}^{(1)}, C_{i,j}^{(2)}, \dots, C_{i,j}^{(k)} \right\}
\end{equation}

Alternatively, this can be expressed in nested form as:
\begin{equation}
\mathcal{S} = \left\{ \left\{ \left\{ C_{i,j}^{(c)} \mid c \in \mathcal{C} \right\} \mid j \in \mathcal{E} \right\} \mid i \in \mathcal{D} \right\}
\end{equation}

This unified structure, instantiated as $\mathcal{S} = \mathcal{QNN}$ or $\mathcal{S} = \mathcal{QF}$, forms the foundation for attention-guided inference, allowing the AI Agent to selectively access clusters based on temporal, hierarchical, or task-specific signals.

\subsubsection{Automated Analytical Pipeline}

Building upon the structured representation $\mathcal{S}$, the AI Agent processes the clustering results through a multi-stage analysis pipeline:

\begin{enumerate}
    \item \textbf{Cluster Meaning Generation:} Derives semantic interpretations for each cluster $C_{i,j}^{(c)}$.
    \item \textbf{Intra-Strategy Summarization:} Aggregates findings within QF and QNN strategies, analyzing intra-epoch patterns.
    \item \textbf{Inter-Strategy Comparison:} Contrasts QF and QNN cluster characteristics under equivalent $K$ values.
\end{enumerate}

This process is formalized in Algorithm~\ref{alg:analytical_pipeline}.

\begin{algorithm}[ht]
\caption{Automated Analytical Pipeline for AI Agent}
\label{alg:analytical_pipeline}
\begin{algorithmic}[1]
\Require Clustered data $\mathcal{S} = \{ \mathcal{S}_i \}$ with $\mathcal{S}_i = \{ \mathcal{E}_{i,j} \}$ and $\mathcal{E}_{i,j} = \{ C_{i,j}^{(c)} \}$
\Statex \textbf{Ensure:} Comparative insights and strategy-level interpretations stored in the Knowledge Base

\For{each depth level $i \in \mathcal{D}$}
    \For{each epoch $j \in \mathcal{E}$}
        \State \mbox{Step 1: Cluster Understanding}
        \For{each cluster $C_{i,j}^{(c)}$}
            \State Generate semantic representation using the \emph{Cluster Meaning Generator}
            \State Store $\phi(C_{i,j}^{(c)})$ into $\texttt{ClusterInfo}(i, j, c)$
        \EndFor
        \State \mbox{Step 2: Intra-Epoch Analysis}
        \State Form cluster set $\texttt{EpochComparison}_{i,j} = \{ C_{i,j}^{(c)} \mid c \in \mathcal{C} \}$
        \State Analyze intra-epoch similarity and partitioning rationale
        \State \mbox{Step 3: Inter-Strategy Comparison (QNN vs. QF)}
        \State Generate QNN-specific summary for $\mathcal{E}_{i,j}$
        \State Generate QF-specific summary for $\mathcal{E}_{i,j}$
        \State Compare QNN and QF clusters and store differences into the Knowledge Base
    \EndFor
\EndFor
\State \mbox{Step 4: Global Strategy Analysis}
\For{each cluster number $k \in \mathcal{C}$}
    \State Compare QNN and QF behavior across $k$
    \State Store global insights into \texttt{StrategyKnowledge}
\EndFor
\Return Updated Knowledge Base
\end{algorithmic}
\end{algorithm}

Notable observations include:
\begin{itemize}
    \item \textbf{Singleton Clusters:} Frequently observed in QNN outcomes, representing extreme transaction behaviors.
    \item \textbf{Diverse Clusters:} Common in QF results, capturing varied token types and address interactions.
\end{itemize}

Finally, the AI Agent synthesizes a global comparison across all $k$ values, summarizing the trade-offs between QNN and QF strategies trade-offs.

\subsubsection{Knowledge Base Structure}

The Knowledge Base retains insights at three levels:
\begin{enumerate}
    \item \textbf{Cluster-Level:} Semantic summaries of individual clusters.
    \item \textbf{Intra-Epoch Comparison:} Contrastive analysis across clusters within the same epoch.
    \item \textbf{Strategy-Level Insights:} Comparative understanding of QF and QNN strategies across all configurations.
\end{enumerate}

Table~\ref{tab:module-kb} summarizes the data flow across modules and the Knowledge Base.

\begin{table*}[!htbp]
\centering
\caption{Module-to-Knowledge Base Data Flow Correspondence Table}
\label{tab:module-kb}
\begin{tabular}{M{3cm} M{4cm} M{3cm} M{4cm} M{3cm}}
\toprule
\multirow{2}{*}{\textbf{Module Name}} 
 & \multicolumn{2}{c}{\textbf{Input}} 
 & \multicolumn{2}{c}{\textbf{Output}} \\
\cmidrule(lr){2-3} \cmidrule(lr){4-5}
 & \textbf{Data} & \textbf{Data Source Level} 
 & \textbf{Data} & \textbf{Data Source Level} \\
\midrule
Cluster Meaning Generation 
 & Each cluster \(C_{i,j}^{(c)}\) (raw cluster information) 
 & Raw block-level 
 & Semantic text description
 & Cluster-level information \\
\midrule
QNN Cluster Summarization 
 & Semantic text generated from Cluster Meaning Generation 
 & Cluster-level information 
 & Cluster summary text produced by the QNN strategy 
 & Intra-epoch comparison \\
\midrule
QF Cluster Summarization  
 & Semantic text generated from Cluster Meaning Generation 
 & Cluster-level information 
 & Cluster summary text produced by the QF strategy 
 & Intra-epoch comparison \\
\midrule
Final Findings Synthesis  
 & QNN and QF cluster summaries and Intra-epoch comparisons results 
 & Intra-epoch comparison 
 & Comprehensive strategic analysis result (final text description) 
 & Strategy-level insight \\
\bottomrule
\end{tabular}
\end{table*}

\begin{table*}[!htbp]
\centering
\caption{Comparison of Best Clustering Performance: Monte Carlo (Quantum Features) vs QNN for $K=2$ to $K=6$}
\begin{tabular}{ccccccc}
\toprule
$K$ & Method & Depth & Epoch & Silhouette Score & Davies-Bouldin & Calinski-Harabasz \\
\midrule
3 & Quantum Features (Worst Run) & 1 & -  & 0.394566 & 0.927151 & 3,847 \\
2 & Quantum Features (Average) & 1 & -  & 0.660680 & 0.470618 & 133,976 \\
2 & Quantum Features (Best Run) & 1 & -  & 0.996368 & 0.042230 & 1,260,727 \\
2 & QNN              & 2 & 1  & \textbf{0.999777}        & \textbf{1.111477e-8}        & \textbf{15,833,657,123,341}         \\
\midrule
3 & Quantum Features (Worst Run) & 1 & -  & 0.383393 & 0.868310 & 4,066 \\
3 & Quantum Features (Average) & 1 & -  & 0.612198 & 0.553668 & 16,307,927,154 \\
3 & Quantum Features (Best Run) & 1 & -  & \textbf{0.999994} & \textbf{0.000798} & 456,616,095,717 \\
3 & QNN                          & 4 & 5 & 0.999510        & 0.141294       & \textbf{168,665,013,271,780}           \\
\midrule
4 & Quantum Features (Worst Run) & 1 & -  & 0.351097 & 0.935585 & 3,891 \\
4 & Quantum Features (Average) & 1 & -  & 0.600428 & 0.577474 & 45,602,215,148 \\
4 & Quantum Features (Best Run) & 1 & -  & 0.998005 & 0.223911 & 1,276,852,339,920 \\
4 & QNN                          & 9 & 9  & \textbf{0.999246}       & \textbf{0.000054}        & \textbf{1,307,035,577,888,410}           \\
\midrule
5 & Quantum Features (Worst Run) & 1 & -  & 0.315429 & 1.058288 & 3,711 \\
5 & Quantum Features (Average) & 1 & -  & 0.600791 & 0.567613 & 61,140,087,483 \\
5 & Quantum Features (Best Run) & 1 & -  & 0.997542 & 0.305401 & 1,711,909,181,014 \\
5 & QNN                          & 6 & 1  & \textbf{0.998789}        & \textbf{0.138369}        & \textbf{7,704,660,919,690,170}           \\
\midrule
6 & Quantum Features (Worst Run) & 1 & -  & 0.328886 & 0.993762 & 3,561 \\
6 & Quantum Features (Average) & 1 & -  & 0.597184 & 0.580199 & 84,564,962,473 \\
6 & Quantum Features (Best Run) & 1 & -  & 0.997905 & 0.262240 & 2,367,802,555,468 \\
6 & QNN                          & 6 & 1  & \textbf{0.998540}        & \textbf{0.000125}        & \textbf{23,337,479,575,752,000}           \\
\bottomrule
\end{tabular}
\label{tab:best_results}
\end{table*}

\subsubsection{Integration with Quantitative Results}

Our two-stage framework tightly couples numerical rigor with qualitative depth. The AI Agent validates and contextualizes the optimal cluster count (typically $K = 3$), providing semantic explanations that align with numerical findings, thereby supporting transparent and explainable decision-making in blockchain analytics.

\begin{enumerate}
    \item \textbf{Cluster-level information:} Each cluster $C_{i,j}^{(c)}$ stores local representations, such as representative transactions, learned embeddings, or feature vectors. Formally:
    \begin{equation}
        \text{ClusterInfo}(i,j,c) = \phi\left( C_{i,j}^{(c)} \right)
    \end{equation}
    where $\phi(\cdot)$ is an embedding or summarization function.
    
    \item \textbf{Intra-epoch comparison:} To support contrastive reasoning among peer clusters within the same depth $i$ and epoch $j$, we define a set:
    \begin{equation}
        \text{EpochComparison}_{i,j} = \left\{ C_{i,j}^{(c)} \mid c \in \mathcal{C} \right\}
    \end{equation}
    which can be used to derive similarity metrics or decision boundaries among clusters at the same time step.
    
    \item \textbf{Strategy-level insight:} At the highest abstraction level, the Knowledge Base retains strategy-level patterns extracted from both QNN and QF representations:
    \begin{equation}
        \text{StrategyKnowledge} = \Psi\left( \mathcal{QNN}, \mathcal{QF} \right)
    \end{equation}
    where $\Psi(\cdot)$ is a strategy integration or contrastive analysis function used to compare or fuse information across models.
\end{enumerate}

Together, these three levels of memory allow the AI Agent to reason within a single cluster and across clusters and feature extraction paradigms, forming the foundation for multi-level decision-making.

\section{Results}

This section presents the results of our two-stage analysis, following the framework outlined in the Methodology section. The evaluation combines quantitative metrics and AI Agent-assisted qualitative insights to comprehensively assess Quantum Features (QF) and fully trained Quantum Neural Networks (QNN) clustering performance. Special attention is given to the effectiveness of a three-cluster ($K=3$) configuration, as identified through our empirical analysis.

\subsection{Stage 1: Quantitative Evaluation Results}

Our experimental results demonstrate that both the Random QNN and the fully trained QNN models outperform classical clustering baselines, with the fully trained QNN achieving the best overall performance. The fully trained QNN attains the highest Silhouette Score of $0.9995$ and the lowest Davies-Bouldin index of $0.0001$, indicating superior cluster compactness and separation.

As summarized in Table~\ref{tab:best_results}, the QNN approach consistently outperforms the Monte Carlo-based quantum random feature method across all evaluated cluster sizes ($K=2$ to $K=6$). Although the quantum random features method demonstrates competitive performance in its best runs, achieving Silhouette Scores exceeding $0.99$ and respectable Calinski-Harabasz values, the fully trained QNN reliably delivers superior clustering quality across all metrics.

For instance, at $K=2$, the QNN achieves a near-perfect Silhouette Score of $0.999777$, an exceptionally low Davies-Bouldin index of $1.11e-8$, and an extraordinary Calinski-Harabasz index surpassing $15$ trillion. This performance advantage persists as clusters increase, with QNN maintaining consistently higher Silhouette Scores and Calinski-Harabasz indices while keeping the Davies-Bouldin index remarkably low. Notably, even at higher $K$ values, the increasing circuit depth and number of training epochs in QNN reflect its adaptability to complex clustering tasks.

These findings highlight the robustness and scalability of the QNN-based approach in quantum-enhanced clustering scenarios. By leveraging parameterized quantum circuits and end-to-end training, the fully trained QNN consistently demonstrates an advantage over Monte Carlo-based quantum feature extraction methods.

\begin{table*}[t]
\centering
\caption{Consolidated Summary of AI Agent-Derived Insights: QNN vs. QF Clustering Strategies for Different K Values}

\begin{tabular}{M{1cm} M{5cm} M{5cm} M{5cm}}
\toprule
\textbf{K} & \textbf{QNN Key Characteristics} & \textbf{QF Key Characteristics} & \textbf{Overall Recommendation} \\
\midrule
2 & 
It focuses on strong individual signals (e.g., large ETH transfer, dominant sender) and produces homogeneous clusters when high values are detected. 
& 
Emphasizes less popular tokens and repeated address interactions. Often produces a "catch-all" cluster with diverse transactions.
& 
QNN effectively isolates large transactions, while QF reveals subtle trading patterns. The choice depends on the analysis goals. \\
\midrule
3 & 
It creates clusters based on clear, individual features (e.g., high ETH and SHIB transfers). This may result in singleton clusters.
& 
Group transactions using a combination of factors (token diversity, temporal aspects, address interactions) for more nuanced clusters.
& 
Use QNN for distinctly defined clusters; opt for QF to capture more complex interrelations. \\
\midrule
4 & 
Aggregates transactions by standard senders and high values, but sometimes isolates single high-value transactions.
& 
It produces a noise cluster while other clusters capture recurring tokens and address patterns.
& 
Neither approach is ideal alone; consider increasing K or tuning parameters for more balanced clustering. \\
\midrule
5 & 
Overly sensitive to transaction value, resulting in singleton clusters and issues with data sparsity.
& 
Provides more meaningful and interpretable clusters by capturing token-specific activity and recurring address interactions.
& 
For the analyzed dataset, QF offers more robust insights than QNN. \\
\midrule
6 & 
Strongly emphasizes extreme values, often forming singleton clusters for high-value transactions.
& 
Focuses on token and relational features, yielding clusters with multiple transactions and consistent sender-receiver patterns.
& 
QNN is suitable for anomaly detection, whereas QF better captures overall transaction patterns. \\
\bottomrule
\end{tabular}
\end{table*}

\subsection{Stage 2: AI Agent-Assisted Qualitative Analysis}

Following the execution of the proposed pipeline, the AI Agent produced LLM-generated outputs that facilitate a qualitative comparison of clustering strategies. These outputs compare the results of the Quantum Neural Network (QNN) and Quantum Feature-based (QF) approaches across various values of $K$. The comparisons address key aspects such as cluster definitions, differences and similarities in transaction grouping, potential reasons for observed discrepancies, and evaluations of the pros and cons for each method.

For instance, at $K=2$, the AI Agent analysis reveals that QNN isolates high-value transactions or single-sender activities. In contrast, QF emphasizes combinations of less popular tokens and recurring address interactions, resulting in more heterogeneous clusters. As $K$ increases, QNN continues producing clusters dominated by high-value transactions or specific address-driven patterns. Conversely, QF clusters consistently capture multifaceted relationships involving token diversity, temporal dynamics, and network flow.

The contrast becomes more apparent for higher values of $K$ ($K=5$ and $K=6$). While QNN remains strongly value-sensitive, isolating substantial transfers and focusing on key addresses, QF continues incorporating token-specific features and recurrent sender-receiver relationships. Table~\ref{tab:clustering-summary} consolidates these insights across different clustering counts.

\begin{table*}[t]
\centering
\caption{Consolidated Summary of AI Agent-Derived Insights: QNN vs. QF Clustering Strategies for Different K Values}
\label{tab:clustering-summary}
\begin{tabular}{M{1cm} M{5cm} M{5cm} M{5cm}}
\toprule
\textbf{K} & \textbf{QNN Key Characteristics} & \textbf{QF Key Characteristics} & \textbf{Overall Recommendation} \\
\midrule
2 & Isolates large transactions and dominant sender patterns. & Captures token diversity and repeated address interactions. & Use QNN for anomaly detection; QF for transactional variety. \\
\midrule
3 & Creates distinct clusters with clear features, including singleton clusters. & Groups transactions using multiple factors for nuanced clusters. & QNN for sharply defined clusters; QF for complex interrelations. \\
\midrule
4 & Aggregates standard senders and high-value transactions, occasionally isolates outliers. & Includes noise clusters alongside recurring patterns. & Increase $K$ or tune parameters for balanced clustering. \\
\midrule
5 & Highly sensitive to transaction value, leading to singleton clusters. & Captures token-specific activities and recurring address interactions. & QF offers more robust insights for this dataset. \\
\midrule
6 & Emphasizes extreme values with frequent singleton clusters. & Focuses on token and relational features for consistent clusters. & QNN for anomaly detection; QF for broader transaction patterns. \\
\bottomrule
\end{tabular}
\end{table*}

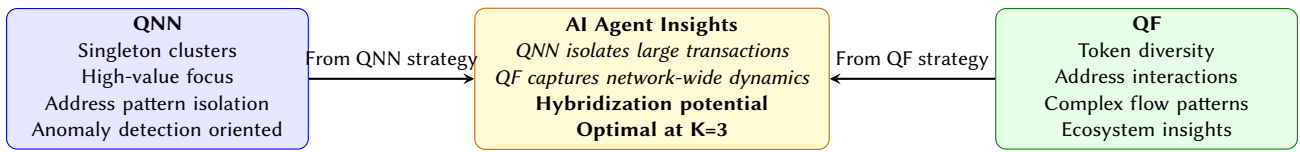
\begin{figure*}[!t]
    \centering
    \begin{tikzpicture}[scale=1, transform shape, 
        node distance=1.5cm and 2.2cm,
        every node/.style={font=\sffamily\small},
        qnnbox/.style={rectangle, draw=blue!80!black, rounded corners, minimum width=3.8cm, minimum height=1.4cm, text width=3.8cm, text centered, align=center, fill=blue!10},
        qfbox/.style={rectangle, draw=green!50!black, rounded corners, minimum width=3.8cm, minimum height=1.4cm, text width=3.8cm, text centered, align=center, fill=green!10},
        middle/.style={rectangle, draw=orange!80!black, rounded corners, minimum width=4.5cm, minimum height=1.8cm, text width=4.5cm, text centered, align=center, fill=yellow!20},
        bottomlabel/.style={font=\bfseries, text width=1\textwidth, align=center},
        arrow/.style={thick, ->, >=stealth}
    ]

    \node[qnnbox] (qnn) {\textbf{QNN} \\ Singleton clusters \\ High-value focus \\ Address pattern isolation \\ Anomaly detection oriented};

    \node[middle, right=of qnn] (insight) {\textbf{AI Agent Insights} \\ \emph{QNN isolates large transactions} \\ \emph{QF captures network-wide dynamics} \\ \textbf{Hybridization potential} \\ \textbf{Optimal at K=3}};

    \node[qfbox, right=of insight] (qf) {\textbf{QF} \\ Token diversity \\ Address interactions \\ Complex flow patterns \\ Ecosystem insights};

\node[fit=(qnn)(qf), inner sep=0pt] (allboxes) {};

    \draw[arrow] (qnn) -- node[midway, above] {From QNN strategy} (insight);
    \draw[arrow] (qf) -- node[midway, above] {From QF strategy} (insight);

    \end{tikzpicture}
    \caption[AI Agent-enhanced semantic comparison of QF and QNN clustering]{AI Agent-enhanced semantic comparison of QF and QNN clustering at $K=3$. QNN emphasizes anomaly detection in high-impact transactions, while QF captures the broader dynamics of token interactions and behavioral diversity within the ecosystem.}
    \Description{AI Agent-enhanced semantic comparison of QF and QNN clustering at K=3, highlighting QNN's focus on anomaly detection and QF's emphasis on token interaction dynamics.}
    \label{fig:qualitative-analysis}
\end{figure*}

As shown in Figure~\ref{fig:qualitative-analysis}, the AI Agent confirms that QNN is particularly effective for detecting high-value anomalies, while QF uncovers broader ecosystem patterns, emphasizing token diversity and behavioral variation. Integrating quantitative and qualitative findings, our analysis converges on $K=3$ as the optimal choice, where clusters balance clarity and complexity, capturing global transaction trends and localized anomalies without excessive fragmentation.

\section{Discussion}

Our findings validate the efficacy of the proposed two-stage framework. The quantitative evaluation underscores the strength of QNN in delivering superior clustering metrics, achieving high Silhouette Scores and Calinski-Harabasz indices, while maintaining low Davies-Bouldin values. These results indicate QNN's effectiveness in constructing well-separated and dense clusters, especially in detecting anomalies within high-impact blockchain transactions.

Complementing the quantitative assessment, the AI Agent provided crucial qualitative insights that deepen our understanding of clustering behaviors. As visualized in Figure~\ref{fig:qualitative-analysis}, QNN's strength lies in its precision — isolating singleton clusters and spotlighting significant transactional patterns- advantageous for anomaly detection and identifying dominant actors. Conversely, QF reveals a broader spectrum of ecosystem activities, capturing token diversity, recurring address interactions, and complex transactional flows, making it particularly valuable for ecosystem-wide behavioral analysis.

This juxtaposition highlights an essential trade-off: while QNN excels at narrowing in on impactful outliers, it risks over-fragmentation and may overlook more subtle patterns. In contrast, QF provides a holistic view of the transaction landscape but may generalize patterns at the expense of pinpointing critical anomalies.

Crucially, our integrated analysis consistently converges on $K=3$ as the optimal clustering configuration. At this granularity, QNN and QF demonstrate complementary strengths: QNN delivers sharp, focused clusters for anomaly detection, while QF uncovers the interconnected dynamics of token relationships and behavioral diversity. The AI Agent's interpretive layer enriches this understanding, confirming that $K=3$ balances complexity and clarity, avoiding over-fragmentation while preserving interpretability.

These insights underscore the potential of adopting a hybrid strategy that leverages both QNN and QF approaches. Such an ensemble framework could dynamically allocate analytical focus, using QNN for high-risk monitoring and QF for comprehensive ecosystem mapping. Ultimately, the synergy between quantitative clustering performance and AI-assisted qualitative interpretation exemplifies a path toward more explainable and actionable blockchain analytics frameworks.

\section{Conclusion}
In this study, we presented a comprehensive two-stage framework combining quantum-enhanced clustering techniques with AI Agent-assisted qualitative analysis to improve the interpretability and performance of blockchain transaction clustering. Leveraging both Random Quantum Features (QF) and fully trained Quantum Neural Networks (QNN), our approach systematically evaluated clustering configurations across multiple cluster counts ($K = 2$ to $K = 6$), providing both quantitative performance metrics and human-interpretable insights.

Our quantitative experiments demonstrated that fully trained QNN models consistently outperformed Monte Carlo-based quantum random features. The QNN achieved superior clustering quality, notably high Silhouette Scores and Calinski-Harabasz indices, while maintaining remarkably low Davies-Bouldin values across all tested configurations. These findings confirm the robustness and scalability of QNNs in unsupervised learning scenarios, especially for complex, high-dimensional blockchain data.

Complementing these quantitative results, the AI Agent performed an in-depth qualitative assessment of clustering outputs. It revealed that QNN clusters tend to isolate high-value transactions and single-actor patterns, making them well-suited for anomaly detection and monitoring high-impact entities. In contrast, QF clusters captured a broader spectrum of transactional behaviors, including token diversity, address interactions, and complex flow patterns, offering richer insights into the overall ecosystem dynamics.

The convergence of quantitative metrics and AI Agent analyses identified $K=3$ as the optimal cluster configuration. This configuration balances clarity and complexity, providing actionable insights while avoiding over-fragmentation. The AI Agent further highlighted the complementary nature of QNN and QF strategies, recommending hybridization to maximize analytical depth and flexibility.

Our research contributes to the emerging field of explainable quantum machine learning by demonstrating how AI Agents can bridge the gap between advanced quantum computation and human interpretability. This integrated approach enhances the transparency of quantum-assisted clustering models and offers practical tools for decision support in blockchain governance, fraud detection, and ecosystem analysis.

Future work will explore dynamic and adaptive clustering frameworks where AI Agents can iteratively refine clustering parameters based on evolving data streams. We plan to extend this methodology to multi-modal blockchain datasets and investigate ensemble models that synergize QNN and QF strengths for greater interpretability and performance.

Our findings underscore the promise of combining quantum machine learning with AI-driven explainability, paving the way for more transparent, robust, and insightful analytics in complex decentralized environments.

\bibliographystyle{ACM-Reference-Format}
\bibliography{sample-base}

\end{document}